\documentclass[prl,twocolumn,showpacs,amsmath,amssymb,superscriptaddress]{revtex4}
\usepackage{graphicx}

\newcommand{\GF}{\ensuremath{\mathcal{G}}} 
\newcommand{\abs}[1]{\ensuremath{\left| #1 \right|}} 

\begin{document}

\title{Nonequilibrium functional renormalization group for interacting
  quantum systems}
\author{Severin~G.~Jakobs}
\affiliation{Institut f\"ur Theoretische Physik~A, RWTH Aachen, D-52056
  Aachen, Germany}
\author{Volker~Meden}
\affiliation{Institut f\"ur Theoretische Physik, Universit\"at G\"ottingen,
  D-37077 G\"ottingen, Germany}
\author{Herbert~Schoeller}
\affiliation{Institut f\"ur Theoretische Physik~A, RWTH Aachen, D-52056
  Aachen, Germany}
\date{\today}

\begin{abstract}
  We propose a nonequilibrium version of functional renormalization 
  within the Keldysh formalism by introducing a 
  complex valued flow parameter in the Fermi or Bose functions of each 
  reservoir. Our cutoff scheme provides a unified approach to
  equilibrium and nonequilibrium situations.
  We apply it to nonequilibrium transport through an interacting quantum
  wire coupled to two reservoirs and show that the nonequilibrium occupation 
  induces new power law exponents for the conductance.
\end{abstract}


\maketitle

{\bf Introduction.}
Relevant energy scales of interacting quantum many-body systems
are typically spread over several orders of magnitude. Conventional
perturbative approaches to such systems which treat all energy scales
at once often fail, especially in low dimensions where they typically
lead to divergences at low energies (see e.g. \cite{solyom}). A
successful approach to these systems is given by 
renormalization group (RG) schemes which are based on the idea
of treating energy scales successively from high to low
\cite{wilson,solyom}. A broadly applicable example is the functional RG (FRG).
In this approach an infrared cutoff is introduced in the bare propagator 
and an exact infinite hierarchy of flow equations for Green or vertex functions 
is derived \cite{salmhofer}. This method benefits from various advantages: 
(i) it is not restricted to field theories but can directly be applied
to microscopic models; (ii) the physics on all energy scales can be studied, 
not only the low energy asymptotics; (iii) it provides a systematic scheme for
approximations allowing for the study of systems with many interacting degrees of 
freedom (``large'' systems). In thermal equilibrium the FRG 
has been used successfully for the study of microscopic models of correlated 
fermions in 0, 1, and 2 dimensions \cite{minireview,metzner}. 
In particular the FRG scheme for the one-particle irreducible 
vertex functions \cite{wetterich,morris,salmhofer_honerkamp}
has the advantage that the self-energy is fed back nonperturbatively
into the FRG flow, which is essential for 0 and 1 dimensional systems
\cite{minireview}.

The FRG is usually formulated within the equilibrium imaginary time
Matsubara formalism. A generalization to the real time Keldysh
approach is highly desirable as it makes nonequilibrium systems
treatable and does not require an elaborate analytic continuation of the
Green functions (GFs). RG methods developed so far in the framework of 
Keldysh formalism \cite{schoeller_koenig, rosch} 
are restricted to systems with only a few interacting degrees of freedom. 
Wegner's flow equation method provides an alternative ``RG-like'' approach to 
nonequilibrium, but up to now was also only used for small  
systems \cite{Kehrein}. 
Recent attempts to develop RG-schemes within Keldysh formalism
(for the steady state) by using a {\it real} frequency or momentum
cutoff in the free propagator  \cite{dipl, gezzi, millis} 
are promising but face many severe technical problems. 
In this Letter we propose a fundamentally new approach
by introducing  an {\it imaginary} frequency cutoff in the
Fermi or Bose functions of the reservoirs, which are coupled to the
quantum system under consideration. If reservoir $r$ is characterized 
by temperature $T_r$ and chemical potential $\mu_r$, we introduce a
multidimensional (or complex-valued) flow parameter $\Lambda =
(\{\Lambda_r\}_r,\lambda)$ by the replacement 
\begin{gather}
  \nonumber 
  f_r(\omega) 
  = 
  \frac{1}{e^{(\omega - \mu_r)/T_r} \pm 1}
  =
  T_r \sum_{\omega_n^r} \frac{e^{i \omega_n^r 0^+}}{i 
    \omega_n^r - \omega +
    \mu_r}
  \\
  \label{eq:L_fermi}
  \rightarrow \quad
  f^{\lambda,\Lambda_r}_r(\omega) 
  =
  T_r \sum_{\omega_n^r} \frac{\Theta(\abs{\omega_n^r} -
    \lambda) e^{i \omega_n^r 0^+}}{i \omega_n^r - \omega + \Lambda_r}
\end{gather}
with $\omega_n^r=(2n+1)\pi T_r$ or $\omega_n^r=2n\pi T_r$
denoting the Matsubara frequencies for fermions or bosons,
respectively.
The imaginary component $\lambda$ of the flow parameter corresponds
to the imaginary frequency cutoff used in Matsubara formalism FRG. For
a simple truncation scheme we show below [see Eq.~\eqref{eq:se_flow}],
that this cutoff results in the identical flow equation as found by
the Mastubara FRG in the special case of equilibrium. For Fermions,
$\lambda$ regularizes infrared divergences by broadening the step in
the Fermi function in a way comparable to an augmented temperature $T
\sim \lambda$. Therefore the flow generated by $\lambda$ is quite
similar to the temperature-flow scheme known from equilibrium
\cite{T_flow}. 
The real components $\Lambda_r$ of the flow parameter correspond to
the chemical potentials of the reservoirs and allow for a continuous
interpolation between equilibrium and nonequilibrium, generalizing 
the known temperature flow scheme
to a voltage-flow scheme for nonequilibrium problems. 

Applying our scheme to quantum transport through a wire coupled
to reservoirs, we show that the nonequilibrium distribution
can change the power law exponent of the local density of states
leading to novel exponents in the conductance. This exemplifies that
nonequilibrium can even modify the {\it scaling behavior}, whereas
recently discussed decoherence effects induced by relaxation processes
associated with the finite current lead to an additional cutoff of
the RG flow \cite{rosch1}. 

{\bf Model and cutoff scheme.}
We consider an interacting quantum
system with Hamiltonian $H_\text{S}=H_\text{S}^{(0)} + V$, where
$H_\text{S}^{(0)}=\sum_l \epsilon_l c^\dagger_l c_l$ is quadratic in
the field operators (including arbitrary impurity configurations),
and $V$ denotes the interaction. This system is coupled to several
noninteracting reservoirs (labeled by the index $r$) described by
$H_{\text{res}}=\sum_r H_{\text{res}}^r = \sum_{kr}
\epsilon_{kr}a^\dagger_{kr} a_{kr}$, which are in a grandcanonical 
distribution with (possibly different) temperatures $T_r$ and chemical 
potentials $\mu_r$. The coupling between $H_\text{S}$ and $H_{\text{res}}$ 
can involve an arbitrary number of system or reservoir field operators,
in order to describe particle and energy exchange or nonlinear
couplings. Although our general formalism is applicable to all these
cases, for simplicity we restrict ourselves in this Letter to the
special case of one kind of fermionic or bosonic particle which can 
tunnel between the reservoirs and the quantum system, described by the tunneling 
Hamiltonian $H_\text{T} = \sum_{klr} t_{kl}^r a^\dagger_{kr} c_l +
\text{H.c}$. We assume interactions which conserve
particle number and consider only the nonequilibrium steady state
limit where time-translational invariance holds.
Given $H= H_\text{S}^{(0)} + H_{\text{res}} + H_\text{T} + V$
and expanding in $H_\text{T}$ and $V$, 
one can apply standard quantum field theoretical diagrammatic 
approaches in imaginary (Matsubara formalism) or real (Keldysh
formalism) times to calculate all (non)equilibrium GFs
from which physical observables like, e.g., charge, spin, or heat currents 
can be deduced. 

The FRG within Matsubara formalism is based on introducing 
an infrared cutoff $\Lambda$ in the free one-particle GF
$g_{\text{eq},l}(i\omega_n)=1/(i\omega_n-\epsilon_l+\mu)$ 
of the quantum system. A choice often used in 0 and 1 dimension \cite{minireview}
is to cutoff the imaginary Matsubara frequencies via 
$g^\Lambda_\text{eq} (i \omega_n) =
\Theta(\abs{\omega_n} - \Lambda) g_\text{eq}(i \omega_n)$.
In the nonequilibrium case, a natural adaption to Keldysh formalism is a 
frequency cutoff in the free Keldysh GF of the quantum system
via $g^\Lambda(\omega) = \Theta(\abs{\omega} - \Lambda)
g(\omega)$. Here $g$ consists of the four components
$g^c, g^<, g^>, g^{\tilde c}$ or the three independent components
$g^\text{Ret}, g^<, g^\text{Av}$. While this approach was used to study 
transport properties of interacting systems in nonequilibrium 
\cite{dipl,gezzi}, it suffers from certain drawbacks when the
truncation schemes developed in equilibrium \cite{minireview} are
applied: 
First the causality relation $\Sigma^c + \Sigma^{\tilde c} = \Sigma^< +
\Sigma^>$ of the self-energy is violated. Second, it was found that 
for stronger interactions singularities in the flow equations are
induced making them numerically difficult to tackle \cite{gezzi}.
Third, the FRG flow along the real axis has to pass the poles of the free
resolvent making a numerical treatment very expensive for large
systems with many discrete single-particle states \cite{dipl}. 
The first (and possibly also the second) 
problem can be circumvented by choosing a cutoff
in momentum instead of frequency space \cite{millis}. This provides a
meaningful approach for homogeneous (bulk) materials, but not for the 
structured, mesosocopic systems we are mainly interested in. In such
setups momentum is not a good quantum number. We note that the third
problem is present for all cutoff schemes on the real frequency 
or momentum axis.

For these reasons, we propose the use of a complex valued flow
parameter with imaginary part $\lambda$ cutting off the Matsubara
poles of the Fermi or Bose functions of the reservoirs and real parts
$\Lambda_r$ describing flowing chemical potentials, as defined by
Eq.~(\ref{eq:L_fermi}).   
It is essential to implement the cutoff in the distribution functions
of the reservoirs and not in the initial distribution function of the
system because the former determine the nonequilibrium occupation of
the system in the steady state whereas the latter does not affect
the steady state at all. 
The distribution functions of the reservoirs appear
diagrammatically in the free lesser GF of reservoir $r$, given by 
$g_{r}^<(\omega) = \mp f_r(\omega)(g_{r}^\text{Ret} -
g_{r}^\text{Av})(\omega)$ 
with $g_{kr}^{\text{Ret}/\text{Av}}(\omega)=1/(\omega-\epsilon_{kr}\pm
i\eta)$. As the insertion of the cutoff is merely a manipulation of
the particle distribution, the causality relation is not violated.

A typical starting point of the FRG flow will be at
$\lambda=\infty$ (with arbitrary $\Lambda_r$) which yields
$f_r^\Lambda \equiv 0$  corresponding to an empty system with known
properties. In particular any possible infrared divergence will be
regularized at this point.
The FRG flow stops at $(\{\Lambda_r =
\mu_r\}_r,\lambda=0)$,  fully restoring all distribution
functions. The path along which the FRG flows through  the
$(\{\Lambda_r\}_r,\lambda)$-space can be chosen in an arbitrary way
without affecting the result at the end of the flow as long as the
infinite hierarchy of equations is treated exactly. In the common case
of truncations however, different paths may lead to different
approximations for the vertex functions. 
Particular paths are defined by letting first flow $\lambda$ to $0$
while keeping the $\Lambda_r$ constant and in a second step adjusting
the $\Lambda_r$ to $\mu_r$. In the model discussed at the end of this
Letter we find that the numerical solution is most stable if one sets
$\Lambda_r=\mu_r$ right from the beginning. 

{\bf Derivation of the flow equations.}
The flow equations for the irreducible vertex functions can be derived
by a simple diagrammatic procedure \cite{dipl} in contrast to the
formalism based on generating functionals \cite{gezzi}.
The $k$-particle vertex function $\gamma_k$ is defined as the sum of
all one-particle irreducible diagrams with $k$ amputated incoming
lines and $k$ amputated outgoing lines. An example is the self-energy
$\Sigma \equiv \gamma_1$. A diagram $D_k$ contributing to $\gamma_k$
consists of vertices representing the interaction $V$ which are
connected by directed lines representing the reservoir dressed GF
$\GF^\Lambda = \left[g^{-1} - \Sigma^\Lambda_\text{res}\right]^{-1}$
of the system. The lines depend on $\Lambda$ via the reservoir
self-energy $\Sigma^\Lambda_\text{res} = \sum_r \Sigma_r^\Lambda$, with
$\Sigma_r^\Lambda(\omega) = (t^r)^\dagger g_r^\Lambda(\omega) t^r$ (in
matrix notation) where the free propagator $g_r^\Lambda$ of the
reservoir contains $f_r^\Lambda$ in its $<$-component.
An irreducible diagram $D_k = D_k^\Lambda$ depends on $\Lambda$ via
its internal lines $\GF^\Lambda$ which enter its value
multiplicatively. Consequently the derivative $\partial_\Lambda
D^\Lambda_k$ is given by product rule as sum over all
diagrams ${D'}^\Lambda_k$ which are identical to $D^\Lambda_k$ except
for having one differentiated GF $\partial_\Lambda \GF^\Lambda$ among
its lines. We represent $\partial_\Lambda \GF^\Lambda$
diagrammatically by a crossed out line. In order to determine
$\partial_\Lambda \gamma^\Lambda_k = \sum \partial_\Lambda
D^\Lambda_k$ we thus have to calculate the sum off all diagrams
${D'}^\Lambda_k$ which result from crossing out any line in any
diagram $D^\Lambda_k$ contributing to $\gamma^\Lambda_k$.   

\begin{figure}
  \begin{center}
    \includegraphics[width = 0.8\linewidth]{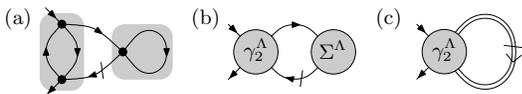}
  \end{center}
  \caption{
    (a) A diagram ${D'}^\Lambda_1$ contributing to
    $\partial_\Lambda \Sigma^\Lambda \equiv \partial_\Lambda
    \gamma_1^\Lambda$. The one-particle irreducible subdiagrams
    connected to a ring are marked by gray underlays.
    (b) When adding up all rings with the same structure, the
    subdiagrams sum to vertex functions.
    (c) Resumming the self-energy insertions yields
    the propagator $S^\Lambda$.
  }
  \label{rings}
\end{figure}

Imagine amputating temporarily the crossed out line from a diagram
${D'}^\Lambda_k$. As ${D'}^\Lambda_k$ was irreducible, the remaining
part is still connected and of such structure that it becomes
irreducible when the crossed out line is reinstalled. This structure is
given by a chain of irreducible subdiagrams connected by single lines
which becomes a closed ring when the crossed out line is reinstalled.
Indeed, except for multiplicity (which cancels certain symmetry
prefactors), the diagrams ${D'}^\Lambda_k$ are exactly all possible rings 
of irreducible subdiagrams closed by a crossed out line.
For an example see Fig.~\ref{rings}(a). In order to determine
$\partial_\Lambda \gamma^\Lambda_k = \sum {D'}^\Lambda_k$ let us first
add up all diagrams ${D'}^\Lambda_k$ which have identical ring
structure and differ only in the form of the irreducible
subdiagrams. The summation of all possible subdiagrams then yields
vertex functions connected to a ring, see Fig.~\ref{rings}(b). By
further summation of rings of vertex-functions which are identical
except for self-energy insertions we obtain full GFs $G^\Lambda =
\GF^\Lambda + \GF^\Lambda \Sigma^\Lambda \GF^\Lambda + \dots$
connecting the vertex functions and one propagator $S^\Lambda =
G^\Lambda {\GF^\Lambda}^{-1} [\partial_\Lambda \GF^\Lambda]
{\GF^\Lambda}^{-1} G^\Lambda $ closing the ring; compare
Fig.~\ref{rings}(c) (where the resulting ring contains only one vertex
function). 

Concluding, $\frac{\partial}{\partial \Lambda} \gamma^\Lambda_k
(\nu'_1 \dots \nu'_k|\nu_1 \dots \nu_k)$ (with incoming indices $\nu_1,
\dots, \nu_k$ and outgoing indices $\nu'_1, \dots, \nu'_k$, where
$\nu$ stand for frequency, state and contour index)
is determined as follows: Draw all rings consisting of vertex functions
$\gamma^\Lambda_2, \gamma^\Lambda_3, \dots, \gamma^\Lambda_{k+1}$
which have the external indices $\nu'_1, \dots, \nu_k$
all in all and which are connected to rings by full propagator lines and
exactly one full crossed out propagator representing $S^\Lambda$.
Evaluate these diagrams as usual, treating the vertex functions as
normal vertices. Because of the special ring structure and the
distinguished position of the crossed out line there is no need for
prefactors corresponding to symmetries or equivalent lines.

$S^\Lambda$ has vanishing retarded and advanced components but 
nonzero lesser-component  
$S^{\Lambda, <}= G^{\Lambda, \text{Ret}} (\partial_\Lambda
\Sigma_\text{res}^{\Lambda, <})  G^{\Lambda, \text{Av}}$ with
$\partial_\Lambda\Sigma_\text{res}^{\Lambda, <}=
\pm i \sum_r (\partial_\Lambda f_r^\Lambda)\Gamma_r$
and $\Gamma_r=i(\Sigma_r^{\text{Ret}}-\Sigma_r^{\text{Av}})$.
In $\lambda$-direction due to 
\begin{equation}
  \label{eq:fermi_dev}
  \frac{\partial}{\partial\lambda} f_r^\Lambda(\omega) = -T_r \sum_{\omega_n^r}
  \frac{\delta(\abs{\omega_n^r} - \lambda)}{i \omega_n^r - \omega + \Lambda_r},
\end{equation}
at most one Matsubara frequency contributes to $S^\Lambda$. This feature
largely reduces the computational effort for a numerical evaluation of
the flow along the $\lambda$-direction and is the reason for calling
$S^\Lambda$ single scale propagator in the Matsubara FRG.
The flow along the $\Lambda_r$-directions benefits
from a similar advantage if the $T_r$ are small. Consider, e.g., the
case $\lambda = 0$. Then $\partial_{\Lambda_r} f_r^\Lambda(\omega)
|_{\lambda = 0} = - \partial_\omega f_r(\omega) |_{\mu_r = \Lambda_r}$
as a function of energy has a peak at $\omega=\Lambda_r$ 
of width $\sim T_r$. Frequency integration has to be carried out only
around this peak, reducing the computational effort.

According to the diagrammatic rules formulated above, the single
contribution to the flow of the self-energy
$\partial_\Lambda\Sigma^\Lambda$  stems from the diagram in Fig.\ \ref{rings}(c)
yielding 
\begin{equation}
  \label{eq:se_flow}
  \frac{\partial}{\partial \Lambda} \Sigma^\Lambda_{\nu' \nu} =
  \mp \frac{i}{2\pi} \sum_{\nu_1,\nu_1^\prime}
  \gamma^\Lambda_2(\nu',\nu_1^\prime|\nu,\nu_1)
  \, S^\Lambda_{\nu_1\nu_1^\prime}\quad.
\end{equation}
In the special case of equilibrium and neglecting the renormalization of 
the 2-particle vertex, it is straightforward to show that this RG
equation is identical to the one of the Matsubara FRG
\cite{meden_metzner, tilman}.

\begin{figure}
  \begin{center}
    \includegraphics[width = 0.75\linewidth]{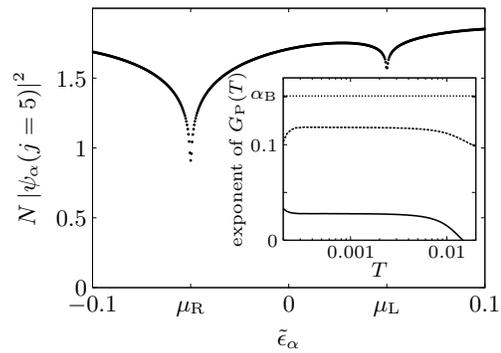}
  \end{center}
  \caption{
    Suppression of the one-particle wave function as function
    of energy near the boundary (at site $5$) at
    $T = 10^{-4}$. Inset: Exponent of the conductance through a probe
    lead weakly coupled to site $5$ as function of $T$
    for $\mu_\text{P}=\mu_\text{L}=0.05$ (solid line) and
    $\mu_\text{P}=\mu_\text{R}=-0.05$ (dashed line).
    Using the same truncation scheme in equilibrium ($\mu_\text{L} =
    \mu_\text{R}=0$), the functional RG yields the
    boundary exponent $\alpha_\text{B} \simeq 0.151$ (dotted line).
    $N=20000$, $U=0.5$, $\tau_\text{L}=0.075$, $\tau_\text{R}=0.15$. 
  }
  \label{fig:result}
\end{figure}

{\bf Quantum wire with contact barriers.} 
As an application of our method we discuss nonequilibrium 
steady state transport through an interacting quantum wire (Luttinger
liquid, LL), end-contacted through tunnel barriers to two
noninteracting reservoirs. In equilibrium strong barriers 
induce slowly decaying ($\sim$ inverse distance) 
oscillatory effective potentials of range 
proportional to the thermal length $\frac{v_F}{T}$ in the LL which
lead to a one-particle spectral
weight which goes as 
$\rho(\omega) \sim \max\{\abs{\omega - \mu},T\}^{\alpha_\text{B}}$
close to the barrier \cite{kane_fisher}. The boundary exponent is given by
$\alpha_\text{B} = \frac{1}{K} - 1$, where the LL parameter $K$ depends
on the strength of the interaction and the filling.
For a long wire contacted via high tunnel barriers by noninteracting
leads this induces a power law for
the linear conductance $G \sim T^{\alpha_\text{B}}$ \cite{tilman}. 
We show that finite bias voltage and asymmetric coupling to
the leads give rise to two different exponents. We use a tight-binding
model for the noninteracting
part $H_0=H_\text{S}^{(0)}+H_{\text{res}} + H_\text{T}$ 
\begin{equation}
  H_0 = -  \sum_{\substack{j \in \mathbb{Z} \\ j \neq 0, N}}
  c^\dagger_{j+1}c^{\phantom{\dagger}}_j - \tau_\text{L} c^\dagger_1
  c^{\phantom{\dagger}}_0 - \tau_\text{R} c^\dagger_{N+1}
  c^{\phantom{\dagger}}_N + \text{H.c.} 
\end{equation}
and take $V=U\sum_{j=1}^{N-1}\rho_j\rho_{j+1}$ 
for the interaction, with $\rho_j=c_j^\dagger c_j - \frac{1}{2}$.
The semi-infinite noninteracting parts $j<1$ and $j>N$ of the chain
represent the left (L) and right (R) reservoir.
We choose $\Lambda_\text{L,R}\equiv \mu_\text{L,R}$ and let $\lambda$ 
flow from $\infty$ to $0$. With $f^{\lambda = \infty} \equiv 0$ the 
flow of the interacting part of the self-energy starts at 
$\Sigma^{\lambda = \infty, \text{Ret}}_{ij}= 0$.
We truncate the hierarchy of flow equations by setting
$\gamma_2^\lambda \simeq \gamma_2^{\lambda = \infty} = \overline{v}$,
an approximation which was used successfully in 
equilibrium \cite{meden_metzner} 
and is justified if the interaction is sufficiently small to neglect the inelastic 
processes it generates. 
In equilibrium our truncated flow equation is
identical to that of the Matsubara FRG and hence known to produce
results  which agree qualitatively to those obtained by
other methods for interactions such that $1/2 < K \leq
1$. Quantitative agreement is achieved for $|1-K| \ll 1$. In
particular, the scaling exponent comes out correctly in 
leading order in $U$~\cite{meden_metzner, tilman}.
For small tunneling $\tau_r\ll 1$, the flow equation is given by
\begin{equation}
  \label{eq:se_flow_neq}
  \frac{\partial \Sigma^{\lambda,\text{Ret}}_{ij}}{\partial \lambda}
  =
  - \sum_{\substack{r,\omega_n^r\\k,l}} T_r \frac{\Gamma_r}{\Gamma}
  \delta(\abs{\omega_n^r} - \lambda)
  \overline{v}_{i k, j l} 
  G_{lk}(i \omega_n^r + \mu_r)
\end{equation}
with $\Gamma = \Gamma_\text{L} + \Gamma_ \text{R}$,
$\Gamma_\text{L}=\Gamma_\text{L}(E)_{11}$ and
$\Gamma_\text{R}=\Gamma_\text{R}(E)_{NN}$ [the energy dependence of $\Gamma(E)$ can
be neglected near the Fermi levels of the reservoirs]. 
Thus the contribution of each reservoir is weighted by $\Gamma_r$,
arising from the fact that the noninteracting 
occupation of states in the wire is given by $[\Gamma_\text{L}
f_\text{L}(\epsilon) + \Gamma_\text{R} f_\text{R}(\epsilon)]/\Gamma$. 
During the FRG flow, each reservoir contribution generates
oscillations of the effective potential which influence the other's
formation via back coupling through Eq.~\eqref{eq:se_flow_neq}. This mutual
influence can be neglected for small interaction leading to two
independent oscillations with period given by the Fermi wavelength and
amplitude proportional to $\Gamma_r$ of the corresponding reservoir.
We define in the single-particle 
subspace an effective Hamiltonian of the
quantum wire by 
$h_{\text{eff}}=H_\text{S}^{(0)}+\Sigma^{\lambda=0,\text{Ret}}$
with eigenfunctions $\psi_\alpha$ and eigenvalues
$\tilde{\epsilon}_\alpha$.
The area of the discrete peak at energy
$E\sim\tilde{\epsilon}_\alpha$ of the local spectral density
$\rho(E)_{jj}$ at site $j$ is proportional to $\abs{\psi_\alpha(j)}^2$,
which scales as $\max\{\abs{\tilde{\epsilon}_\alpha - \mu_\text{L}},
T_L\}^{\alpha_\text{L}}  
\max \{\abs{\tilde{\epsilon}_\alpha -
\mu_\text{R}},T_R\}^{\alpha_\text{R}}$ if $j$ is close to the
boundary \cite{remark_Gamma}, where $\alpha_r =
\alpha_\text{B}(\mu_r)\Gamma_r/\Gamma$; see 
Fig.~\ref{fig:result}. Experimentally, this can be 
tested by attaching weakly a probe lead with chemical potential
$\mu_\text{P}$ to the wire near the left
or right boundary at position $j_\text{P}$. Using the Landauer-B\"uttiker 
formula we obtain for
$\Gamma_\text{P}\ll\Gamma_\text{L},\Gamma_\text{R}$ and 
$T_\text{P}$ larger than the level spacing that 
the conductance in the third lead is given by
\begin{equation}
  G_\text{P}=e\frac{\partial I_\text{P}}{\partial \mu_\text{P}}
  =-e^2\Gamma_\text{P}
  \sum_\alpha \abs{\psi_\alpha(j_\text{P})}^2
  f_\text{P}'(\tilde{\epsilon}_\alpha).
\end{equation}
Thus the conductance $G_\text{P}$ scales as $T^{\alpha_r}$ for
$\mu_\text{P}=\mu_r$, with $r=L,R$, as shown in the inset of
Fig.~\ref{fig:result}. To obtain this result we used a particular path
in the complex ``cutoff space''. The exponents remain unchanged if a
different path is chosen. Similar results were obtained applying
bosonization to an effective field theory \cite{Chud}.

{\bf Summary.}
We have shown that the FRG can be generalized in a natural and numerically
efficient and stable way to study nonequilibrium transport through
interacting systems by introducing a complex flow parameter to the
distribution function of each reservoir. We explained how the flow
equations of the FRG can be understood in terms of diagrams and set up
diagrammatic rules to determine the flow equations. We applied
our method to a microscopic model for steady state transport through
an end-contacted, interacting quantum wire and
found that nonequilibrium  changes the scaling exponents by making
them dependant on the height of the contact barriers.

We thank T.~Korb, Th.~Pruschke, F.~Reininghaus, A.~Rosch, and
K.~Sch\"onhammer for helpful discussions. This work was supported by
the DFG via the Forschergruppe 723 and the VW Foundation and the
Forschungszentrum J\"ulich via the virtual 
institute IFMIT (S.G.J. and H.S.). 



\begin{thebibliography}{99}

\bibitem{solyom}
  J.~S\'olyom, Adv.~Phys.~\textbf{28}, 201 (1979).

\bibitem{wilson}
  K.~G.~Wilson and J.~Kogut, Phys.~Rep.~\textbf{12}, 75 (1974).

\bibitem{salmhofer}
  M.~Salmhofer, \textit{Renormalization: An Introduction} (Springer, Heidelberg, 1998). 

\bibitem{minireview} V.~Meden, in
\textit{Advances in Solid State Physics}, Vol.~46, edited by R.~Haug
(Springer, New~York, 2007); arXiv:cond-mat/0604302.

\bibitem{metzner}
  W.~Metzner, Prog.~Theor.~Phys.~Suppl.~\textbf{160}, 58 (2005).

\bibitem{wetterich}
  C.~Wetterich, Phys.~Lett.~B~\textbf{301}, 90 (1993).

\bibitem{morris}
  T.~R.~Morris, Int.~J.~Mod.~Phys.~A~\textbf{9}, 2411 (1994). 

\bibitem{salmhofer_honerkamp}
  M.~Salmhofer and C.~Honerkamp, Prog.\ Theor.\ Phys.\ \textbf{105}, 1 (2001).

\bibitem{schoeller_koenig}
  H.~Schoeller and J.~K\"onig, Phys.~Rev.~Lett.~\textbf{84}, 3686 (2000).

\bibitem{rosch}
  A.~Rosch \textit{et al.},
  Phys.~Rev.~Lett.~\textbf{90}, 076804 (2003). 

\bibitem{Kehrein} 
  S.~Kehrein, Phys.~Rev.~Lett.~\textbf{95}, 056602 (2005).  

\bibitem{dipl}
  S.~Jakobs, Diploma thesis, RWTH Aachen (2003).
  
\bibitem{gezzi}
  R.~Gezzi, Th.~Pruschke, and V.~Meden, Phys.\ Rev.\ B~\textbf{75}, 045324
  (2007).

\bibitem{millis}
  A.~Mitra \textit{et al.},
  Phys.~Rev.~Lett.~\textbf{ 97}, 236808 (2006).

\bibitem{T_flow}
C.~Honerkamp and M.~Salmhofer, Phys.~Rev.~Lett.~{\bf 87}, 187004 (2001). 

\bibitem{rosch1}
  A.~Rosch, J.~Kroha, and P.~W\"olfle,
  Phys.~Rev.~Lett.~\textbf{87}, 156802 (2001).

\bibitem{meden_metzner}
  V.~Meden \textit{et al.}, 
  J.~Low~Temp.~Phys. \textbf{126}, 1147 (2002).

\bibitem{tilman}
  T.~Enss \textit{et al.},
  Phys.~Rev.~B~\textbf{71}, 155401 (2005).

\bibitem{kane_fisher}
  C.~L.~Kane and M.~P.~A.~Fisher, 
  Phys.~Rev.~B~\textbf{46}, 15233 (1992).

\bibitem{remark_Gamma}
  More precisely, $\abs{\psi_\alpha(j)}^2$ scales as
  $\max\{\abs{\tilde{\epsilon}_\alpha - \mu_\text{L}},
  T_L, \Gamma_{\alpha \alpha}\}^{\alpha_\text{L}} \max
  \{\abs{\tilde{\epsilon}_\alpha -
    \mu_\text{R}},T_R, \Gamma_{\alpha
    \alpha}\}^{\alpha_\text{R}}$. But in the limit of high contact
  barriers, $\Gamma_{\alpha \alpha}$ is much smaller than the level
  spacing and does not contribute visibly to the broadening of the
  power law dips.

\bibitem{Chud} M.~Trushin and A.L.~Chudnovskiy, arXiv:0705.4552. 

\end{thebibliography}
\end{document}